\documentclass[aps,prl, twocolumn,nofootinbib,superscriptaddress]{revtex4}%
\usepackage{color}
\usepackage{epsfig}
\usepackage{graphicx}
\usepackage{amsmath}
\usepackage{amsfonts}
\usepackage{amssymb}
\usepackage{epstopdf}
\usepackage{bbold}%
\providecommand{\U}[1]{\protect\rule{.1in}{.1in}}
\def\be{\begin{equation}}
\def\ee{\end{equation}}
\begin{document}
\title{More on the Matter of 6D SCFTs}
\author{Jonathan J. Heckman\footnote{e-mail: jheckman@email.unc.edu}}

\affiliation{Department of Physics, University of North Carolina, Chapel Hill, NC 27599, USA}
\affiliation{Jefferson Physical Laboratory, Harvard University, Cambridge MA, 02138, USA}

\begin{abstract}
M5-branes probing an ADE singularity lead to 6D SCFTs with $(1,0)$ supersymmetry.
On the tensor branch, the M5-branes specify domain walls of a 7D Super Yang-Mills theory
with gauge group $G$ of ADE-type, thus providing conformal matter for a broad class of generalized quiver theories.
Additionally, these theories have $G \times G$ flavor symmetry, and
a corresponding Higgs branch. In this note we use the F-theory realization of
these theories to calculate the scaling dimension of the operator
parameterizing seven-brane recombination, i.e. motion of the stack of M5-branes off of the orbifold singularity.
In all cases with an interacting fixed point, we find that this operator
has scaling dimension more than six, and defines an irrelevant deformation.
\end{abstract}
\maketitle

\vspace{-3mm}

\section{Introduction}

\vspace{-3mm}

An open question in the study of 6D superconformal field
theories (SCFTs) is to determine the spectrum of operators and their scaling dimensions.
In recent work, a number of new 6D SCFTs have been constructed
by compactifying F-theory on non-compact singular elliptically fibered Calabi-Yau
threefolds. This has already led to a classification of $(1,0)$ theories
without a Higgs branch \cite{Heckman:2013pva}, and has
also been extended to specific theories with a Higgs branch \cite{Heckman:2014} (see also \cite{Gaiotto:2014lca}).
For earlier work on the construction of $(1,0)$ theories see
e.g. \cite{Witten:1995gx, Ganor:1996mu, Seiberg:1996vs, Bershadsky:1996nu, Intriligator:1997kq,
Blum:1997mm, Intriligator:1997dh, Brunner:1997gf, Hanany:1997gh}.
The F-theory realization in particular provides a powerful way to characterize
many aspects of these theories.

In this note we extract the scaling dimension of certain operators in
the $\mathcal{T}(G,N)$ theories of reference \cite{Heckman:2014}. These
theories are given by a stack of $N$ M5-branes probing
an ADE\ singularity $\mathbb{C}^{2}/\Gamma_{G}$ with
$\Gamma_{G}$ a finite ADE subgroup of $SU(2)$. Separating the
M5-branes along the line transverse to this singularity, we see that
each M5-brane specifies a domain wall in the 7D Super Yang-Mills theory defined by
the ADE singularity \cite{Heckman:2014}. Here, the flavor symmetry of the system
is $G_{L}\times G_{R}$, with both factors isomorphic to $G$,
the corresponding ADE-type Lie group.

Our aim will be to extract the dimension of the operator parameterizing motion of the M5-branes off of the
orbifold singularity. In the F-theory realization of these SCFTs, this corresponds to a brane recombination
operation, and we shall refer to it as such. We use the holomorphic geometry of F-theory to extract
data about this branch of the theory.

\textbf{Note Added 07/14/20: The original version of this paper actually performs a 
computation in a 5D Kaluza--Klein (KK) regulated theory. This result 
must then be lifted back to six dimensions to extract the 
corresponding scaling dimensions in 6D SCFTs.}

\vspace{-3mm}

\section{Collisions and Scaling Dimensions}

\vspace{-3mm}

We consider the collision of two singularities in an F-theory
compactification, each supporting an ADE-type gauge group $G$.
At the intersection, we have a superconformal matter sector.
We would like to know the scaling dimension of the operator associated with
Higgsing the flavor symmetry by brane recombination. So, to begin, we consider the local geometries
for such collisions:%
\begin{align}
(E_{8},E_{8})  &  :y^{2}=x^{3}+(uv)^{5}\\
(E_{7},E_{7})  &  :y^{2}=x^{3}+(uv)^{3}x\\
(E_{6},E_{6})  &  :y^{2}=x^{3}+(uv)^{4}\\
(D_{p},D_{p})  &  :y^{2}=(uv)x^{2}+\left(  uv\right)  ^{p-1}\\
(A_{k},A_{k})  &  :y^{2}=x^{2}+(uv)^{k+1},
\end{align}
where $u$ and $v$ are local coordinates of the base. In all cases but the
A-type, the collision leads to a singular geometry requiring further blowups
in the base. Performing such a blowup, we get at least one additional
exceptional curve, which can be wrapped by a D3-brane. When this curve
collapses to zero size, we get a tensionless string.

Now, this string is a BPS\ object, so its tension is given by the exact
formula:
\begin{equation}
\text{tension}=\int_{\Sigma}J,
\end{equation}
with $J$ the K\"ahler form of the Calabi-Yau. On the other hand, the Calabi-Yau geometry tells us that the scaling
of the holomorphic three-form $\Omega$ is related to the K\"{a}hler form via
$J \wedge J \wedge J \sim \Omega\wedge\overline{\Omega}$.
So, we learn that the holomorphic three-form
scales with mass as:
\begin{equation}
\left[  \Omega\right]  \sim\text{tension}^{3/2}\sim\text{mass}^{3}.
\end{equation}
Our plan will be to use this to extract operator scaling
dimensions for our system.

Observe that to really state a relation between the K\"ahler form and the threefold, we need to allow the 
volume modulus of the elliptic curve to be a physical mode. This in turn means that we are really working 
on the F-theory background $S^1 \times CY_3$ and we are actually dealing with a 5D Kaluza--Klein (KK) theory. 
Our answers are therefore obtained with respect to this 5D KK theory, and to extract scaling dimensions in the 
6D SCFT we will need to use the relation:
\begin{equation}\label{5DKK}
\Delta_{6D} = \frac{4}{3} \Delta_{5D,KK},
\end{equation}
where the constant of proportionality follows from inspection of a free hypermultiplet in 6D and 5D.

The argument we will give is well-known in the context of lower-dimensional
systems, and has been used to extract the scaling dimension of the Coulomb
branch parameter for $\mathcal{N}=2$ SCFTs in four dimensions, as in reference
\cite{Argyres:1995xn}. In four dimensions, the homogeneity argument continues to work
for $\mathcal{N}=1$ systems in four dimensions which have a Coulomb branch \cite{Heckman:2010qv},
but the absolute scaling must be determined via a-maximization \cite{Intriligator:2003jj}. Here,
the novelty is that the Calabi-Yau geometry informs us of the Higgs branch,
and that we are extending this analysis to higher-dimensional field theories.

As just mentioned, the next step in our analysis will involve relating the
scaling dimension of the holomorphic three-form back to the scaling dimension
of operators in the SCFT. To this end, let us recall that the brane
recombination operation is controlled by activating vevs for operators. In the
M-theory description, this corresponds to moving the M5-branes off of the orbifold singularity. This breaks
the $G_L \times G_R$ global symmetry down to the diagonal subgroup $G_{diag}$. In all
the cases above, this amounts to the substitution:
\begin{equation}\label{RECCO}
uv\mapsto uv+r.
\end{equation}
One should view the parameter $r$ as the vev of a singlet of $G_{diag}$ which is built from an operator $\mathcal{O}_{rec}$ of the
original CFT. It is natural to expect that just as in the weakly coupled setting, $\mathcal{O}_{rec}$ transforms in the $\mathrm{\mathbf{adj}}_L \otimes \mathrm{\mathbf{adj}}_R$
representation of $G_L \times G_R$. However, determining
this representation is not crucial for our present considerations; All that matters is that the decomposition of
$\mathcal{O}_{rec}$ into irreducible representations of $G_{diag}$ contains a singlet.

Further support for this picture comes from the BPS equations of motion for
the flavor branes, which are controlled by
the Hitchin system coupled to defects \cite{BHVI}:
\begin{equation}
F+\left[  \Phi,\Phi^{\dag}\right]  =\mu_{\mathbb{R}}\delta_{p}\text{ \ \ \ and
\ \ \ }\overline{\partial}_{A}\Phi=\mu_{\mathbb{C}}\delta_{p},
\end{equation}
where $p$ denotes the collision of $u=0$ and $v=0$ in the base. Vevs of operators in the CFT translate to moment maps in the Hitchin system, which
in turn translate to complex structure deformations. This in turn leads to deformations such as the brane recombination operation of line \eqref{RECCO}.

Now, in the configuration of F-theory collisions, it follows from the symmetry of the system that
the coordinates $u$ and $v$ have the same scaling dimension. Further,
we see that $r$, and thus $\mathcal{O}_{rec}$ has twice the scaling dimension of $u$.

But this can be determined directly from the geometry! To see how
to extract this, observe that the holomorphic three-form is given by:%
\begin{equation}
\Omega=\frac{dx}{y}\wedge du\wedge dv.
\end{equation}
So, once we fix the relative scaling dimensions for the coordinates of the threefold,
the absolute scaling of $\Omega$ will allow us to extract the scaling dimension of the
brane recombination operators.

Let us now proceed to the various collisions.
Consider first the geometries:%
\begin{equation}
y^{2}=x^{3}+(uv)^{k}%
\end{equation}
for $k=3,4,5$, which respectively covers $D_{4},E_{6}$ and $E_{8}$. Homogeneity
yields the scaling relations:%
\begin{equation}
y\sim L^{3}\text{, \ \ }x\sim L^{2}\text{, \ \ }r\sim(uv)\sim L^{6/k}\text{,}%
\end{equation}
for some scaling parameter $L$. Relating this back to the scaling of the
holomorphic three-form, we find:
\begin{align}
(E_{8},E_{8}) &  :\dim r_{KK}=18\\
(E_{6},E_{6}) &  :\dim r_{KK}=9\\
(D_{4},D_{4}) &  :\dim r_{KK}=6,
\end{align}
where $\dim r_{KK}$ refers to the scaling dimension in the 5D KK theory. 
By a similar token, for the cases:%
\begin{equation}
y^{2}=x^{3}+(uv)^{l}x,
\end{equation}
we learn, for $l=3$ (i.e. $E_{7}$) and $l=2$ (i.e. $D_{4}$):%
\begin{align}
(E_{7},E_{7}) &  :\dim r_{KK}=12\\
(D_{4},D_{4}) &  :\dim r_{KK}=6,
\end{align}
And in the case of colliding D-type singularities, we get:%
\begin{equation}
(D_{p},D_{p}):\dim r_{KK}=6.
\end{equation}

Finally, in the case of colliding A-type singularities, our method is not really
valid. The reason is that the fiber at the collision
is still in Kodaira-Tate form, so there is nothing to
blow up. This means we have no physical string coming from a D3-brane
wrapped on a collapsing $\mathbb{P}^1$, and consequently, no way to fix the absolute
scaling of the holomorphic three-form. Indeed, there is
not even an interacting fixed point in this case.

Next, we proceed to all of the $\mathcal{T}(G,N)$ theories. This is obtained
by starting with the collision of two $G$-type singularities and performing a
further quotient by $(u,v)\mapsto(\zeta u,\zeta^{-1}v)$ for $\zeta=\exp(2\pi
i/N)$. This leads to an additional singularity at the point $u=v=0$ in the
base of the F-theory geometry. Now, the important point for us is that the
coordinates $u$ and $v$ are no longer valid in the quotient geometry, but
$u^{N}$ and $v^{N}$ are. This means that a full brane
recombination amounts to the substitution:
\begin{equation}\label{reccoN}
uv\mapsto\left(  uv\right)  ^{N}+r_{(N)},
\end{equation}
where $r_{(N)}$ is the vev of our new brane recombination operator. By homogeneity,
a similar scaling analysis now reveals:
\begin{equation}
\dim r_{(N)} = N \cdot \dim r_{(N = 1)}.
\end{equation}

Another way to see this same scaling behavior is to consider a resolution of the $\mathbb{C}^2 / \mathbb{Z}_N$ singularity.
When we do this, we partially move onto the tensor branch of the theory, reaching our generalized quiver theory
with $(G,G)$ conformal matter between each symmetry factor. The singularity resolves to
$N-1$ compact $\mathbb{P}^1$'s, each of which supports a seven-brane with gauge group $G$.
So for each such curve (and the non-compact ones as well), introduce homogeneous coordinates
$[u_i , v_i]$ for $i=1,...,N+1$. The patch $u_i = 1$ indicates the north pole, and $v_i = 1$ indicates the south pole.
Here, $i=1$ (resp. $N+1$) denotes the curve for $G_L$ (resp. $G_R$).
For example, in the case of $E_8$ singularities, each local collision will look like:
\begin{equation}
y^2 = x^3 + (u_i v_{i+1})^5
\end{equation}
for $i=1,...,N$. The local recombination operation for each pair is:
\begin{equation}
u_i v_{i+1} \mapsto u_i v_{i+1} + r_{i,i+1}.
\end{equation}
Performing one such recombination corresponds to moving that particular M5-brane off of the ADE singularity. In the
M-theory picture, the number of domain
walls in the 7D SYM theory defined by the ADE singularity
goes down by one, and in the F-theory picture the
number of compact $\mathbb{P}^{1}$'s goes down by one. The
recombination vev of equation \eqref{reccoN} is now given by
the product $r_{(N)} \sim r_{1,2} \cdots r_{N,N+1}$.
Based on this, it is tempting to view the
aggregate recombination operator as the composite $\mathcal{O}_{(N)} = \mathcal{O}_{1,2} \cdots \mathcal{O}_{N,N+1}$, in the obvious notation.
Note that for $N > 1$, the operator $\mathcal{O}_{i,i+1}$ is not gauge invariant, though its Casimirs will be.

Summarizing, the scaling dimension of the brane recombination operator is given by:
\begin{equation}%
\begin{tabular}
[c]{|c|c|c|c|c|c|}\hline
& $(E_{8},E_{8})$ & $(E_{7},E_{7})$ & $(E_{6},E_{6})$ & $(D_{p},D_{p})$ &
$(A_{k},A_{k})$\\\hline
dim $r_{(N)}$ & $24N$ & $16N$ & $12N$ & $8N$ & $4N$\\\hline
\end{tabular}
\ \ ,
\end{equation}
where here, we have listed the scaling dimensions in the 6D SCFT (see equation \eqref{5DKK}).  
Here, $N \geq 1$ for all entries but the A-type case, where $N \geq 2$. Indeed, as
we already mentioned, we need at least one collapsing $\mathbb{P}^1$ to apply
our scaling argument.

Interestingly, in all cases where our analysis applies, we expect to have an
interacting conformal fixed point in which the scaling
dimension of this operator is greater than six.

\textit{Acknowledgements:} JJH thanks M. Del Zotto, A. Tomasiello, D.R. Morrison and C. Vafa for helpful
discussions and collaboration on related work. JJH also thanks M. Del Zotto, T. Dumitrescu,
A. Tomasiello, and C. Vafa for comments on an earlier draft.
JJH thanks the organizers of the workshop ``Frontiers in String
Phenomenology'' for kind hospitality at Schloss Ringberg during
the completion of this work. The work of JJH was supported in
part by NSF grant PHY-1067976.

\vspace{-3mm}

\end{document}